\documentclass[12pt]{article}
\usepackage{amssymb,amsmath,cite,geometry,graphicx,url}
\catcode`\@=11

\@addtoreset{equation}{section}
\def\theequation{\arabic{section}.\arabic{equation}}
     
\catcode`\@=11
\def\thesection{\arabic{section}.}

\def\appendix{\setcounter{section}{0}
        \def\thesection{Appendix.}
        \def\theequation{\Alph{section}.\arabic{equation}}}
\def\section{\@startsection{section}{1}{\z@}{3.5ex plus 1ex minus
   .2ex}{2.3ex plus .2ex}{\large\bf}}
     
\long\def\@makefntext#1{\parindent 0cm\noindent
\hbox to 1em{\hss$^{\@thefnmark}$}#1}

\newcommand{\captionfonts}{\small}
\makeatletter  
\long\def\@makecaption#1#2{%
  \vskip\abovecaptionskip
  \sbox\@tempboxa{{\captionfonts #1: #2}}%
  \ifdim \wd\@tempboxa >\hsize
    {\captionfonts #1: #2\par}
  \else
    \hbox to\hsize{\hfil\box\@tempboxa\hfil}%
  \fi
  \vskip\belowcaptionskip}
\makeatother   

\begin{document}
\begin{titlepage}
\vspace{.5in}
\begin{flushright}
May 2024\\  
revised July 2024\\
\end{flushright}
\vspace{.5in}
\begin{center}
{\Large\bf
 Causal Sets and an Emerging Continuum}\\  
\vspace{.4in}
{S.~C{\sc arlip}\footnote{\it email: carlip@physics.ucdavis.edu\\
\hspace*{.5em} ORCID 0000-0002-9666-384X }\\
       {\small\it Department of Physics}\\
       {\small\it University of California}\\
       {\small\it Davis, CA 95616}\\{\small\it USA}}
\end{center}

\vspace{.5in}
\begin{center}
{\large\bf Abstract}
\end{center}
\begin{center}
\begin{minipage}{4.5in}
{\small 
Causal set theory offers a simple and elegant picture of discrete physics.  
But the vast majority of causal  sets  look nothing at all like continuum spacetimes, 
and must be excluded in some way to obtain a realistic theory.
I describe recent results showing that almost all non-manifoldlike causal sets are,
in fact, very strongly suppressed in the gravitational path integral.
This does not quite demonstrate the emergence of a continuum---we do not
yet understand the remaining unsuppressed causal sets well enough---but it is
a significant step in that direction.
}
\end{minipage}
\end{center}
\end{titlepage}
\addtocounter{footnote}{-1}

\section{The trouble with causal sets}

Suppose you decide to build a discrete model of spacetime.  There are many
places to start, and your choice may depend on how you view ordinary
continuum spacetimes.  When we learn general relativity, we are usually
taught to think of spacetime as a differentiable manifold $M$, with a topology, 
a Lorentzian signature metric $g$, and perhaps some additional features.  This
naturally suggests discrete structures like lattices, simplicial complexes, 
cell complexes, and the like.

At least for causal manifolds,\footnote{The technical requirement is that
the spacetimes be past and future distinguishing, a slightly weaker condition than 
the absence of closed causal curves.}  though, the same information is contained 
in the causal structure---that is, which points are to the past and future of which 
others---and the local volume element \cite{HKM,Malament}.  This suggests a
very different discrete picture, that of a causal set \cite{Bombelli}.

Loosely speaking, a causal set is a collection of discrete points with prescribed
causal relations, but no other structure.  This naturally fits the requirements of
\cite{HKM,Malament}: the causal structure is given, and the volume of a
region is simply the number of points in contains.  More precisely, a causal
set is a partially ordered set, where $x\prec y$ is interpreted as ``$x$ is to the 
causal past of $y$,'' and the standard antisymmetry condition (``there are no distinct points $x,y$ 
such that $x\prec y$ and $y\prec x$'') becomes a statement of causality.  One further 
postulate is added:\\[1ex]
\hspace*{1em}-- Discreteness: for any $(x,y)$, the set $\{z|x\prec z\prec y\}$ has finitely many
points.\\[1ex]
The set $\{z|x\prec z\prec y\}$, denoted as $(x,y)$, is sometimes called the interval or order interval,  
and is the discrete analog of a causal diamond, the intersection of the future of $x$ and
the past of $y$.  Such intervals will reappear later as building blocks of causal set
invariants.

This is an attractive picture, perhaps the simplest possible version of a discrete spacetime.
Before we get too excited, though, we should remember that the spacetime around us
doesn't look discrete.  So we must first ask how well a causal set---or perhaps a quantum
superposition of causal sets---can approximate a continuum spacetime.

This is best viewed as two distinct questions.  In one direction, if we start with a continuum
spacetime $(M,g)$, can we find a causal set $C$ that approximates it, and use $C$ to
reconstruct the properties of $(M,g)$?  In the other direction---perhaps more important if we
want to consider causal sets as fundamental---if we start with a causal set $C$, can we
find a continuum spacetime $(M,g)$ that approximates it?

In the first direction, we are in reasonably good shape.  There is a standard procedure for 
constructing a causal set from a $d$-dimensional spacetime $(M,g)$, called a ``Poisson sprinkling'' 
\cite{Poisson}: we randomly select a set of points $\{x_i\}$ from $M$ at a fixed density $\rho$ 
with respect to the volume measure $\sqrt{|g|}$; record the causal relations among these
points as determined by the metric $g$; and then ``forget'' $(M,g)$, keeping only the 
points $\{x_i\}$ and their causal relations.  The density $\rho$ determines a discretization scale 
$\ell = \rho^{-1/d}$, and the ``sprinkled'' causal set contains fairly little information about
$M$ at scales smaller than $\ell$.  But above the discretization scale, we understand how
to reconstruct the dimension $d$ \cite{Myrheim,Meyer} (this is less trivial than it might seem), 
the coarse-grained topology and much of the geometry of $M$ (see \cite{Surya} for a
review), as well as useful objects such as d'Alembertians \cite{Benincasa,Glaser,Glaser2} 
and Greens functions (for instance, \cite{Sorkin0,Johnston,Albertini}).  Some of this reconstruction 
is hard---it is particularly difficult to obtain quantities that are local in spacetime, since the only 
intrinsic distance in a causal set is proper distance \cite{Moore,Bombelli2,Glaser3}---and there are 
still important open problems, particularly revolving around the question of how to quantitatively 
measure how close causal sets are to each other or to a manifold.  But there are no obvious
insurmountable difficulties.

 \begin{figure}
\centerline{\includegraphics[width=3.5in]{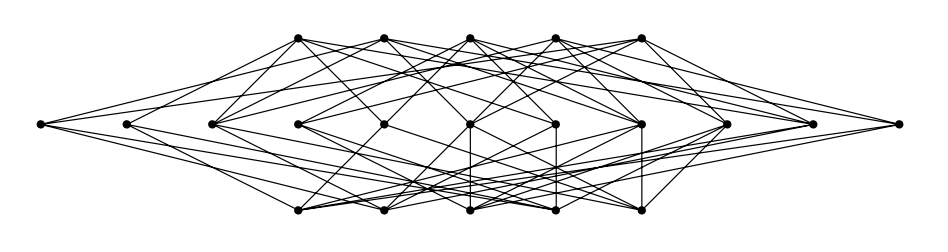}}
\vspace*{-3ex}
\caption{A Kleitman-Rothschild order; lines denote ``links,'' causal nearest neighbors \label{fig1}}
\end{figure}
 The second direction, on the other hand, is a disaster.  Almost all causal sets look nothing
at all like manifolds.  In fact, if you choose a large causal set at random, you will be
overwhelmingly likely to find that you have a Kleitman-Rothschild (or KR) order \cite{KR}.
A KR order is a particular type of ``three-layer'' causal set, that is, a causal set with arbitrarily 
large spatial extent but only three moments of time\footnote{It is tempting to think of 
such a set as a ``thickened'' lower dimensional manifold, but additional restrictions on the 
connectivity rule this out; for instance, the Myrheim-Meyer dimension of a KR order is 
 $\sim 2.38$.}  (see figure \ref{fig1}).  And ``overwhelmingly likely'' here is meant in a 
very strong sense: in the limit that the number of points $n$ goes to infinity, the set of causal
sets that are not KR orders is of measure zero.

This is a strange and unintuitive result, although one may gain some feeling for it by starting
with an $(n-1)$-element KR order and estimating the number of ways of adding an $n$th
point, either in one of the three existing layers or in a fourth layer.  One might hope that this 
non-manifoldlike behavior is just a peculiarity of three-layer sets, but it is not.  If one removes the 
KR orders by hand, almost all remaining causal sets have just two layers, two moments of time.  
If these are removed, almost all of  the remaining causal sets have four layers, then five, six,
and so on, in an extended hierarchy \cite{Dhar,Promel}.  The manifoldlike sets---those that can be 
obtained by a Poisson sprinkling on some spacetime---seem to be a vanishingly small subset.

This does not look good for the causal set program.  Before despairing, though, it is helpful to recall
a similar situation.  If one looks carefully at the standard path integral for a simple system%
---say, a free particle---one finds that almost none of the paths that appear in the sum over
histories are smooth.  In fact, almost all paths are nowhere differentiable \cite{DeWitt}.  This is 
not a problem, though: the contributions of the nonsmooth paths destructively interfere, leaving 
an effectively smooth description.  We might hope that the dominant ``entropic'' contribution
of non-manifoldlike causal sets might be similarly eliminated.  To see whether this is the
case, though, we will need to understand the causal set version of the gravitational path integral. 

\section{The causal set path integral}

Consider a collection $\Omega$ of causal sets---for instance, the collection of all causal
sets with $n$ elements, or the collection of all KR orders.  The partition function for
$\Omega$ is the path integral, or strictly speaking the path sum, 
\begin{align}
Z[\Omega] = \sum_{C\in\Omega}\exp\left\{\frac{i}{\hbar}I[C]\right\}  ,
\label{b1}
\end{align}
where $I$ is a suitable action.  Note the factor of $i$ in the exponent:
causal sets are intrinsically Lorentzian, though there are some ideas of how to continue to
a Euclidean-like sum \cite{Surya2}.

\begin{figure}
\begin{center}
\scalebox{.7}{
\begin{picture}(200,120)(30,0)
\thicklines
\put(20,20){\circle*{3}}
\put(20,50){\circle*{3}}
\put(20,20){\line(0,1){30}}
\put(12,0){$\scriptstyle \in N_0$}
\put(90,20){\circle*{3}}
\put(90,50){\circle*{3}}
\put(90,80){\circle*{3}}
\put(90,20){\line(0,1){60}}
\put(82,0){$\scriptstyle \in N_1$}
\put(160,20){\circle*{3}}
\put(160,50){\circle*{3}}
\put(160,80){\circle*{3}}
\put(160,110){\circle*{3}}
\put(160,20){\line(0,1){90}}
\put(180,0){$\scriptstyle \in  N_2$}
\put(205,35){\circle*{3}}
\put(205,95){\circle*{3}}
\put(185,65){\circle*{3}}
\put(225,65){\circle*{3}}
\put(205,35){\line(2,3){20}}
\put(205,35){\line(-2,3){20}}
\put(205,95){\line(2,-3){20}}
\put(205,95){\line(-2,-3){20}}
 \end{picture}
 }
\end{center}
\vspace*{-2ex}
\caption{$0$-, $1$-, and $2$-element intervals \label{fig2}}
\end{figure}
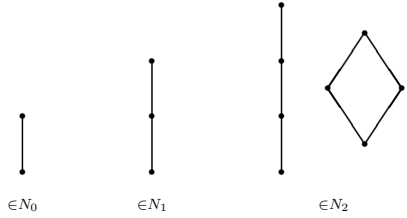

The task is now to find an action that reduces in some sense to the ordinary Einstein-Hilbert
action for the continuum.  For causal sets obtained from a Poisson sprinkling, this task was
accomplished by Benincasa, Dowker, and Glaser \cite{Benincasa,Glaser,Glaser2}, inspired by 
an earlier suggestion of Sorkin \cite{Sorkin}.  The building blocks are causal diamonds, or
order intervals, $(x,y) = \{z|x\prec z\prec y\}$.  An interval can be partially characterized by
the number of elements in its interior, where by convention the initial and final points $x$ and 
$y$ are not counted.  A $0$-element interval is thus a link, a pair of nearest neighbors that are
causally related but have no points lying between them.  Figure \ref{fig2} shows the possible
topologies of $0$-, $1$-, and $2$-element intervals.

For a given causal set $C$, let $N_J(C)$ be the number of intervals in $C$ with exactly $J$ 
elements.  If $C$ is obtained from a sprinkling, the number of elements in an interval is proportional
to its spacetime volume, which in turn depends on the curvature.  Using this fact,
Benincasa, Dowker, and Glaser (BDG) showed that the discrete causal set version of the 
Einstein-Hilbert action in $d$ dimensions takes the form
\begin{align} 
\frac{1}{\hbar} I^{(d)}_{\hbox{\tiny\it BDG}}(C) 
   = -\alpha_d\left( \frac{\ell}{\ell_p}\right)^{d-2}\left( n + \frac{\beta_d}{\alpha_d}
  \sum_{J=0}^{n_d} C_J^{(d)} N_J \right) ,
\label{b2} 
\end{align} 
where  $\ell_p$ is the Planck length (with the convention $\ell_p^{d-2} = 8\pi G\hbar$), $\ell$ 
is the discreteness scale, and $\alpha_d$, $\beta_d$,  and $C_J^{(d)}$ are known 
order one dimension-dependent constants. (See \cite{Glaser2} for closed form expressions 
for these constants; their $i$ is my $J+1$.)  In particular, as the sprinkling density $\rho$
goes to infinity, the BDG action for a region $U$ approaches the Einstein-Hilbert action for $U$.
For causal sets corresponding to manifolds with boundary, we also have a good candidate
for the analog of the Gibbons-Hawking boundary term \cite{Buck}.  

The action (\ref{b2}) was derived for causal sets obtained from 
a Poisson sprinkling.  Let us demand, however, that it be applicable for all causal sets.  Then by 
evaluating the partition function (\ref{b1}) for various classes of non-manifoldlike causal sets, we 
can see whether they are, in fact, suppressed by destructive interference.

\section{Suppression of ``bad'' causal sets}

It is one thing to write down the path sum (\ref{b1}), but quite another thing to evaluate it.
For large $n$, the number of causal sets with $n$ elements grows as $2^{n^2/4}$, and
causal sets are completely classified only up to $n=16$ \cite{Brinkmann}.  Some tricks are 
needed.
\begin{figure}
\centerline{\includegraphics[width=3.5in]{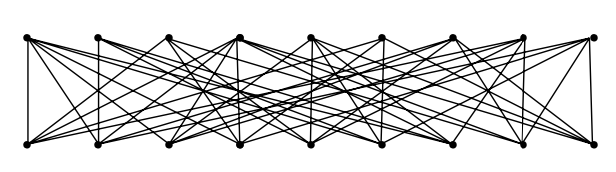}}
\vspace*{-2ex}
\caption{A two-layer causal set \label{fig3}}
\end{figure}

As a first step, in \cite{Loomis} Loomis and I looked at the special case of two-layer sets (see figure
\ref{fig3}), inspired by Loomis's observation that for such sets the BDG action 
becomes drastically simpler.  The simplification occurs because for these sets, 
the only intervals are $0$-element intervals, or links.  Two points can be causally related, 
but with only two layers there is no room for a third point to lie between them.
As a result, the action reduces to the ``link action''
\begin{align}
\frac{1}{\hbar} I^{(d)}_{\hbox{\tiny\it link}}(C) 
   = -\alpha_d\left( \frac{\ell}{\ell_p}\right)^{d-2}\left( n + \frac{\beta_d}{\alpha_d}N_0 \right) ,
\label{c1} 
\end{align} 
where I have used the fact that $C_0^{(d)}=1$ for all dimensions \cite{Glaser2}.  

An $n$-point causal set must have fewer than $N_{\hbox{\tiny\it max}} =  \frac{n^2}{2}$ links, 
and can be characterized by a linking fraction $0\le p\le 1$, where $N_0 = pN_{\hbox{\tiny\it max}}$.  
For a two-layer set, it is easy to see that $p\leq\frac{1}{2}$, with the maximum achieved when the
points are evenly split between the two layers and maximally linked.  In the large $n$ limit, $p$ 
can be treated as a continuous variable, and the path sum (\ref{b1}) for $n$-element two-layer sets 
becomes
\begin{align}
Z[\Omega_{\hbox{\tiny\it n,2-layer}}] =e^{-i\alpha_d\left(\frac{\ell}{\ell_p}\right)^{d-2}n}%
   \int_0^{1/2}\!dp\,|\Omega_{n,2,p}| e^{-\frac{i\beta_d}{2}\left(\frac{\ell}{\ell_p}\right)^{d-2}pn^2} ,
\label{c2}
\end{align}
where $|\Omega_{n,2,p}|$ is the cardinality of the set $\Omega_{n,2,p}$ of two-layer causal 
sets with $n$ elements and linking fraction $p$.  
In \cite{Loomis}, upper and lower bounds for $|\Omega_{n,2,p}|$ were found, which ``squeeze''
its value as $n$ becomes large, yielding
\begin{align}
\ln|\Omega_{n,2,p}| = \frac{1}{4}h(2p) +  {\it o}(n^2) \quad \hbox{with} \quad h(x) = -x\ln x - (1-x)\ln(1-x),
\label{c3}
\end{align}
where $h(x)$ is the entropy function.  The integral (\ref{c2}) can then be evaluated by steepest descents.
Note that a simple saddle point approximation is not enough; one must be careful to choose the correct
saddle point and the correct contour deformation and, since the saddle contribution is exponentially 
small, one must ensure that the contribution from the remainder of the contour is at least as strongly
suppressed.
 
 The result is that for large $n$,
\begin{align}
Z[\Omega_{\hbox{\tiny\it n,2-layer}}] &\sim e^{-bn^2}\nonumber\\
  &\hbox{with} \ b>0 \ \hbox{when} \
  \tan\left\{\frac{\beta_d}{2}\left(\frac{\ell}{\ell_p}\right)^{d-2}\right\} > \left(\frac{27}{4}e^{-1/2} -1\right)^{1/2} .
\label{c4}
\end{align}
For $d=4$, the coefficient $b$ is positive as long as the discreteness scale is $\ell\gtrsim 1.136\ell_p$,
with comparable limits in other dimensions.  We do not know whether these limits are sharp.  Note
that this implies a \emph{very} strong suppression: for a region of $1\,\hbox{cm}^3\times 1\,\hbox{ns}$
with Planck scale discreteness, $n\sim 10^{130}$, so two-layer causal sets are suppressed
by a factor of order $e^{-10^{260}}$.

The calculations leading to (\ref{c4}) were based on the bulk BDG action, and did not include a
Gibbons-Hawking boundary term.  But it may be checked that the boundary terms in \cite{Buck}
are proportional to the cardinality $n$ of the causal set, and contributes only at subleading order.
The same is true for all layered sets; the addition of a boundary term does not affect their
suppression in the path sum.

This is a good first step: the two-layer sets are the second most common ``bad'' causal sets.  But a
direct extension to the KR orders and other layered sets is difficult.  As soon as more than two layers
are present, intervals more complicated than links appear, and the full sum in (\ref{b2}) has to be
accounted for.  In four dimensions, for instance, $n_4=3$, and we must enumerate, or at least
estimate, the numbers $N_J$ with $J=0,1,2,3$.

To make the task controllable, Mathur, Singh, and Surya asked a simpler question \cite{MSS}.
Suppose we truncate the full BDG action (\ref{b2}) by omitting all the
$N_{J>0}$ terms, reducing it to the link action (\ref{c1}).  Then the path sum will again take
the general form (\ref{c2}), but with the combinatoric factor $\Omega_{n,2,p}$ replaced by
the corresponding factor for some other collection of layered causal sets.  The combinatorics
is more complicated, but in the end, the factor in the path sum turns out to be the same as in
the two-layer case, and the suppression (\ref{c4}) reappears for layered sets with more than
two layers.  Note that for this calculation, the dimension $d$ only appears in the factor $\beta_d$
in (\ref{c4}), which gives a very mild dependence of the suppression scale; I will return to this
observation in the conclusion.

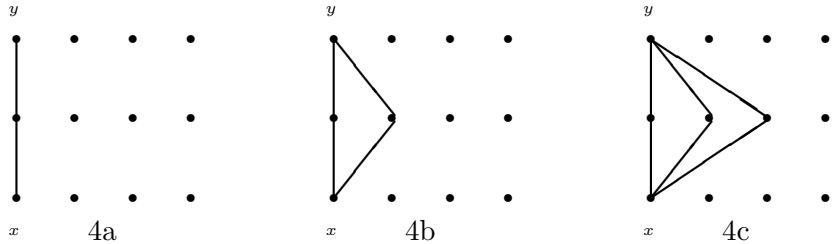
\begin{figure}
\begin{center}
\begin{picture}(320,90)(0,0)
\thicklines
\put(10,20){\circle*{3}}
\put(10,50){\circle*{3}}
\put(10,80){\circle*{3}}
\put(10,20){\line(0,1){60}}
\put(32,20){\circle*{3}}
\put(32,50){\circle*{3}}
\put(32,80){\circle*{3}}
\put(54,20){\circle*{3}}
\put(54,50){\circle*{3}}
\put(54,80){\circle*{3}}
\put(76,20){\circle*{3}}
\put(76,50){\circle*{3}}
\put(76,80){\circle*{3}}
\put(7,6){\tiny $x$}
\put(7,90){\tiny $y$}
\put(37,4){\small 4a}
\put(130,20){\circle*{3}}
\put(130,50){\circle*{3}}
\put(130,80){\circle*{3}}
\put(130,20){\line(0,1){60}}
\put(152,20){\circle*{3}}
\put(152,50){\circle*{3}}
\put(152,80){\circle*{3}}
\put(174,20){\circle*{3}}
\put(174,50){\circle*{3}}
\put(174,80){\circle*{3}}
\put(196,20){\circle*{3}}
\put(196,50){\circle*{3}}
\put(196,80){\circle*{3}}
\put(130,20){\line(4,5){23}}
\put(130,80){\line(4,-5){23}}
\put(127,6){\tiny $x$}
\put(127,90){\tiny $y$}
\put(157,4){\small 4b}
\put(250,20){\circle*{3}}
\put(250,50){\circle*{3}}
\put(250,80){\circle*{3}}
\put(250,20){\line(0,1){60}}
\put(272,20){\circle*{3}}
\put(272,50){\circle*{3}}
\put(272,80){\circle*{3}}
\put(294,20){\circle*{3}}
\put(294,50){\circle*{3}}
\put(294,80){\circle*{3}}
\put(316,20){\circle*{3}}
\put(316,50){\circle*{3}}
\put(316,80){\circle*{3}}
\put(250,20){\line(3,2){44}}
\put(250,80){\line(3,-2){44}}
\put(250,20){\line(4,5){23}}
\put(250,80){\line(4,-5){23}}
\put(247,6){\tiny $x$}
\put(247,90){\tiny $y$}
\put(277,4){\small 4c}
\end{picture}
\end{center}
\vspace*{-2ex}
\caption{One-, two- and three-element intervals in a KR order \label{fig4}}
\end{figure}

This is a good second step, but it is not quite answering the right question.  The link action may
suppress layered causal sets, but it is not a discretization of the Einstein-Hilbert action; for
a Poisson sprinkling, it includes additional nonlocal terms along the light cone \cite{Belenchia}.
The next step, taken in \cite{CCS1}, was to show that for the KR orders the truncation to the
link action is harmless.  More specifically, it was shown that for all KR orders except perhaps a 
set of measure zero, the number $N_{J>0}$ in the full action (\ref{b2}) grow only as $n$, the total 
number of elements of the causal set.  The number $N_0$ of links, in contrast, grows as $n^2$.
Thus for large $n$, the terms omitted in \cite{MSS} give only subleading corrections to the path sum.
While the full proof in \cite{CCS1} requires some careful combinatorics, the intuitive argument 
is fairly simple.  Figure \ref{fig4} shows intervals $(x,y)$ of various sizes in a KR order.  In
figure 4a, a one-element interval contributes to the number $N_1$ in the action.  As soon
as a second path is added between $x$ and $y$, as in figure 4b, $(x,y)$ becomes a 
two-element interval, contributing to $N_2$ but \emph{not} to $N_1$.  A third path, figure 4c,
shifts the contribution of $(x,y)$ to $N_3$.  In general, in order for $(x,y)$ to contribute to
$N_J$, it is not enough that there are $J$ paths from $x$ to $y$; there must be \emph{only}
$J$ paths.  For small causal sets, this is not a terribly strong restriction, but as $n$ becomes
large, the probability of having only a few paths between a point in layer $1$ and a point in
layer $3$ becomes very small, shrinking fast enough to limit the growth of the $N_J$ to 
$\mathcal{O}(n)$.

In a final step, the argument for KR orders has now been generalized to arbitrary 
layered causal sets, as long as the number of layers is small relative to the total number
of points \cite{CCS2}.  A more careful definition of ``layer'' is needed, but it matches the
definition used in \cite{Promel} in the proof of the entropic dominance of layered
sets.  The details of the combinatorics are more complicated than for the KR orders, but the
spirit is the same.  Indeed, given two points $x$ and $y$ in different layers, the addition of more 
layers between them allows for more possible paths, making it even less likely that the interval 
$(x,y)$ contains only a small number of elements.

Combined with the results of \cite{MSS}, this implies that as long as the discreteness length
obeys the inequality (\ref{c4}), all layered causal sets are extremely strongly suppressed in the
ordinary discrete Einstein-Hilbert path sum.  The suppression depends only very mildly on the details
of the action, including the dimension $d$.  It does, however, depend strongly on the
presence of layering.  For causal sets obtained from Poisson sprinklings on Minkowski
space---and very probably on arbitrary curved spacetimes---the $N_J$ all grow as $n^{2-\frac{2}{d}}$
as the number of points increases \cite{Glaser3}, so all of the terms in the BDG action (\ref{b2})
remain important at large $n$.  

The difference can be traced back to the fact that generic layered causal  sets have no spatial 
structure, while sets obtained by sprinkling retain a memory of locality.  For a KR order like
that of figure \ref{fig1}, for instance, a typical causal diamond (or order interval) starting in layer 
$1$ and ending in layer $3$ contains about half of the points in layer $2$ in its interior \cite{KR}, 
while a similar causal diamond obtained from a sprinkling has a much smaller interior.  The BDG
action ``prefers'' this remnant of spatial locality, in ways that we do not yet fully understand.

\section{Are we there yet?}

Recall the dilemma we started with: almost all causal sets have a ``layered'' structure that
is nothing like our observed spacetime.  As we have now seen, the combined results of
\cite{Loomis,MSS,CCS1,CCS2} have solved this problem, showing that for a very wide
range of coupling constants, these non-manifoldlike causal sets are very strongly
suppressed in the path sum.  Given the counting arguments of \cite{KR,Dhar,Promel},
we can say, at least technically correctly, that almost all non-manifoldlike causal sets are 
suppressed in this way.

This is not, however, the end of the story.  We do not yet understand enough about 
the space of causal sets to know what is left.  Manifoldlike causal sets are not eliminated, at 
least not in this fashion, but there may well be other non-manifoldlike sets that survive
as well.  Indeed, we don't even know how to tell that a given causal set is
``manifoldlike''; the term is usually taken to mean ``obtainable from a Poisson sprinkling
of Lorentzian manifold,'' but that does not give us an algorithm we can apply to an arbitrary 
causal set.  One feature we certainly want is emergent locality.  The BDG action seems to 
retain some memory of this property---causal diamonds should not be ``too big'' spatially---but 
this is poorly understood.

Nor do we know how to derive the BDG action itself from first principles.  The action (\ref{b2})
was obtained by starting with the Einstein-Hilbert action on a smooth manifold and
looking for a discretized version on a Poisson sprinkling.  This makes sense if we think of 
causal sets only as approximations to a fundamentally smooth spacetime.  But if we view 
causal sets as fundamental, we need a way to derive an action without reference to a
continuum limit.

As noted earlier, the results described here do not need the full structure of the BDG action.
The suppression of layered sets comes from the link term in the action, and the point of 
\cite{CCS1,CCS2} was to show that the higher order terms were not relevant for layered
sets.  To recover anything like general relativity, though---or even spacetime locality
\cite{Belenchia}---we need those higher order terms, with the right coefficients.  This might
require radical steps, for instance a search for a quantum sequential growth model that
has the correct classical limit \cite{Surya}.  But perhaps a simpler approach is available.  One 
possible avenue would be to formulate a renormalization group flow for  the action; perhaps 
the coefficients in the BDG expression label a fixed point.  For now, though, this is more
a dream than a concrete program.

\vspace{1.5ex}
\begin{flushleft}
\large\bf Acknowledgments
\end{flushleft}

I would like to thank Sam Loomis, Sumati Surya, and Peter Carlip, my collaborators
for almost all of the research summarized here.
This work was supported in part by Department of Energy grant
DE-FG02-91ER40674.

\end{document}